\documentclass[twocolumn,prd,showpacs,preprintnumbers,nofootinbib]{revtex4}
\usepackage{graphicx}
\usepackage{epsfig}
\usepackage{bm}
\usepackage{latexsym,amssymb,amsmath,amsfonts,amssymb,txfonts,pxfonts,wasysym,float}
\usepackage{color}

\DeclareBoldMathCommand\blpar{\left(} 
\DeclareBoldMathCommand\brpar{\right)}

\newcommand{\beq}[1]{\begin{equation}\label{#1}}
\newcommand{\eeq}{\end{equation}}
\newcommand{\bea}[1]{\begin{eqnarray} \label{#1}}
\newcommand{\eea}{\end{eqnarray}}
\newcommand{\ba}{\begin{array}}
\newcommand{\ea}{\end{array}}

\def\be{\begin{equation}}
\def\ee{\end{equation}}
\def\gs{\mathrel{
   \rlap{\raise 0.511ex \hbox{$>$}}{\lower 0.511ex \hbox{$\sim$}}}}
\def\ls{\mathrel{
   \rlap{\raise 0.511ex \hbox{$<$}}{\lower 0.511ex \hbox{$\sim$}}}}

\newcommand{\comment}[1]{}

\usepackage[usenames,dvipsnames]{xcolor}
\definecolor{orange}{cmyk}{0,0.5,1,0}
\definecolor{rossoCP3}{cmyk}{0,.88,.77,.40}
\definecolor{graa}{rgb}{0.8,0.8,0.8}
\definecolor{blaa}{rgb}{0.2,0.2,0.6}

\begin{document}

\title{\color{rossoCP3} Hubble Hullabaloo and String Cosmology}

\author{Luis A. Anchordoqui}
\affiliation{Physics Department, Herbert H. Lehman College and Graduate School, The City University of New York\\ 250 Bedford Park Boulevard West, Bronx, New York 10468-1589, USA}

\date{May 2020} 

\begin{abstract}
 \noindent The discrepancy in measurements of the Hubble constant
 indicates new physics in dark energy, dark matter, or both. Drawing
 inspiration from string theory, we explore possible solutions to
 overcome the $H_0$ problem. We investigate the interplay between the
 cosmological determination of $\Delta N_{\rm eff}$  and $Z'$ searches
 at the LHC Run3. 
\end{abstract}
\maketitle

{\bf \color{rossoCP3} \underline{THE STORYLINE:}} The concordance model of
cosmology, with dark energy ($\Lambda$), cold dark matter (CDM),
baryons, and three flavors of one helicity state neutrinos
(left-handed $\nu_L$ along with their right-handed $\overline \nu_R$)
provides a consistent description of big-bang nucleosynthesis (BBN),
the cosmic microwave background (CMB), and the galaxy formation
epoch. However, despite the impressive successes of $\Lambda$CDM in
describing a wide range of cosmological data, various discrepancies
have persisted. Most strikingly, the emerging tension in the inferred
values of the Hubble constant
$H_0 = 100~h~{\rm km} \, {\rm s}^{-1} \, {\rm Mpc}^{-1}$. $H_0$
parametrizes the expansion rate and thus provides clues about the
cosmological energy content of the universe. Drawing inspiration from
string theory,  we explore possible solutions to overcome the
$H_0$ problem.

\begin{figure}[tb]
\centering
\includegraphics[width=0.97\columnwidth]{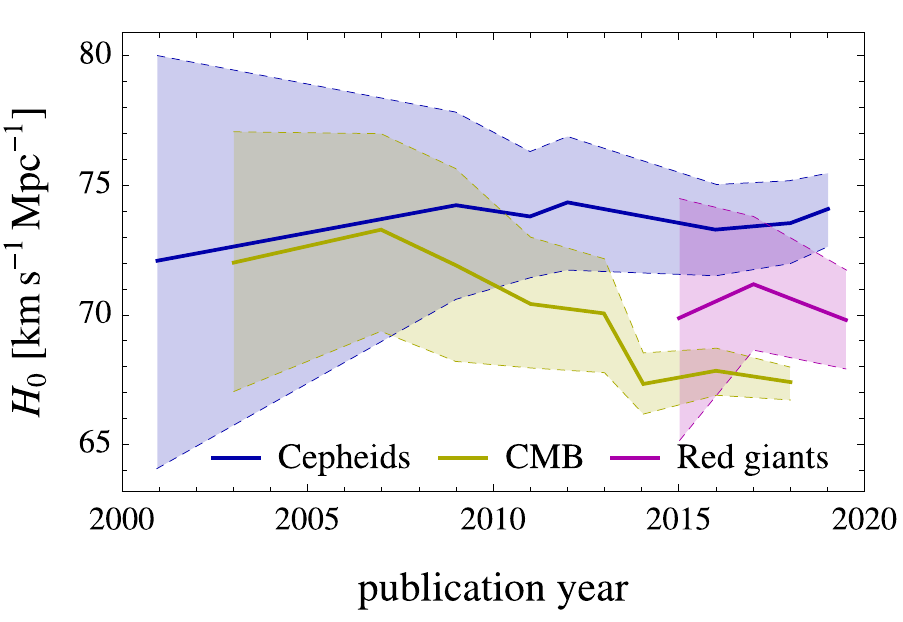} 
    \caption{The Hubble constant as a function of the
      publication
      date. The solid lines indicate the evolution of the mean 
      measurements and the shaded regions span values within one
      standard deviation of the mean. The blue color represent values
      of $H_0$ determined in the nearby universe with a calibration
      based on the Cepheid distance scale  applied to SNe Ia. The first measurement is from the
      Hubble Key Project~\cite{Freedman:2000cf}, the next two measurements
      are from the SH0ES group~\cite{Riess:2009pu,Riess:2011yx},
      the fourth measurement is from the Carnegie Hubble Program which used mid-infrared data to recalibrate the data
from the Hubble Key Project~\cite{Freedman:2012ny}, and the last three measurements are also from
the SH0ES
group~\cite{Riess:2016jrr,Riess:2018byc,Riess:2019cxk}. The brown color
indicates derived values of $H_0$ based on the  $\Lambda$CDM
model and measurements of the CMB. The first five measurements are
from
WMAP~\cite{Spergel:2003cb,Spergel:2006hy,Komatsu:2008hk,Komatsu:2010fb,Hinshaw:2012aka},
the next two are from the Planck mission~\cite{Ade:2013zuv,Ade:2015xua}, then there is the
estimate from the dark energy
survey + baryon acoustic oscillations + BBN~\cite{Abbott:2017smn}, and the last
point is also from the Planck mission~\cite{Aghanim:2018eyx}. The red
color indicates local $H_0$ measurements with a calibration
based on the tip of the red-giant branch distance scale applied
      to SNe
      Ia~\cite{Jang:2015,Jang:2017dxn,Freedman:2019jwv,Freedman:2020dne}. Adapted
      from~\cite{Freedman:2019jwv}.
      \label{fig:1}}
\end{figure}

To set up some context of the current $H_0$ measurements, we start
the discussion with the classical distance ladder approach. This method
combines Cepheid period-luminosity relations with absolute distance
measurements to local anchors so as to
calibrate distances to supernovae type Ia (SNe Ia) host galaxies in the
Hubble flow. In Fig.~\ref{fig:1} we show various values of the Hubble
constant as a function of the publication year. We can see 
that the uncertainties have continued to come down with time. Beginning with the
$H_0$ determination of the Hubble Key Project that has an uncertainty of
roughly 
10\%~\cite{Freedman:2000cf}, the error bars have
reduced considerably to about 1.9\% in the latest result from the
SH0ES group that gives $H_0 = 74.03 \pm 1.42~{\rm km} \, {\rm s}^{-1} \, {\rm Mpc}^{-1}$~\cite{Riess:2019cxk}.

\begin{figure}[tb]
\centering
\includegraphics[width=0.97\columnwidth]{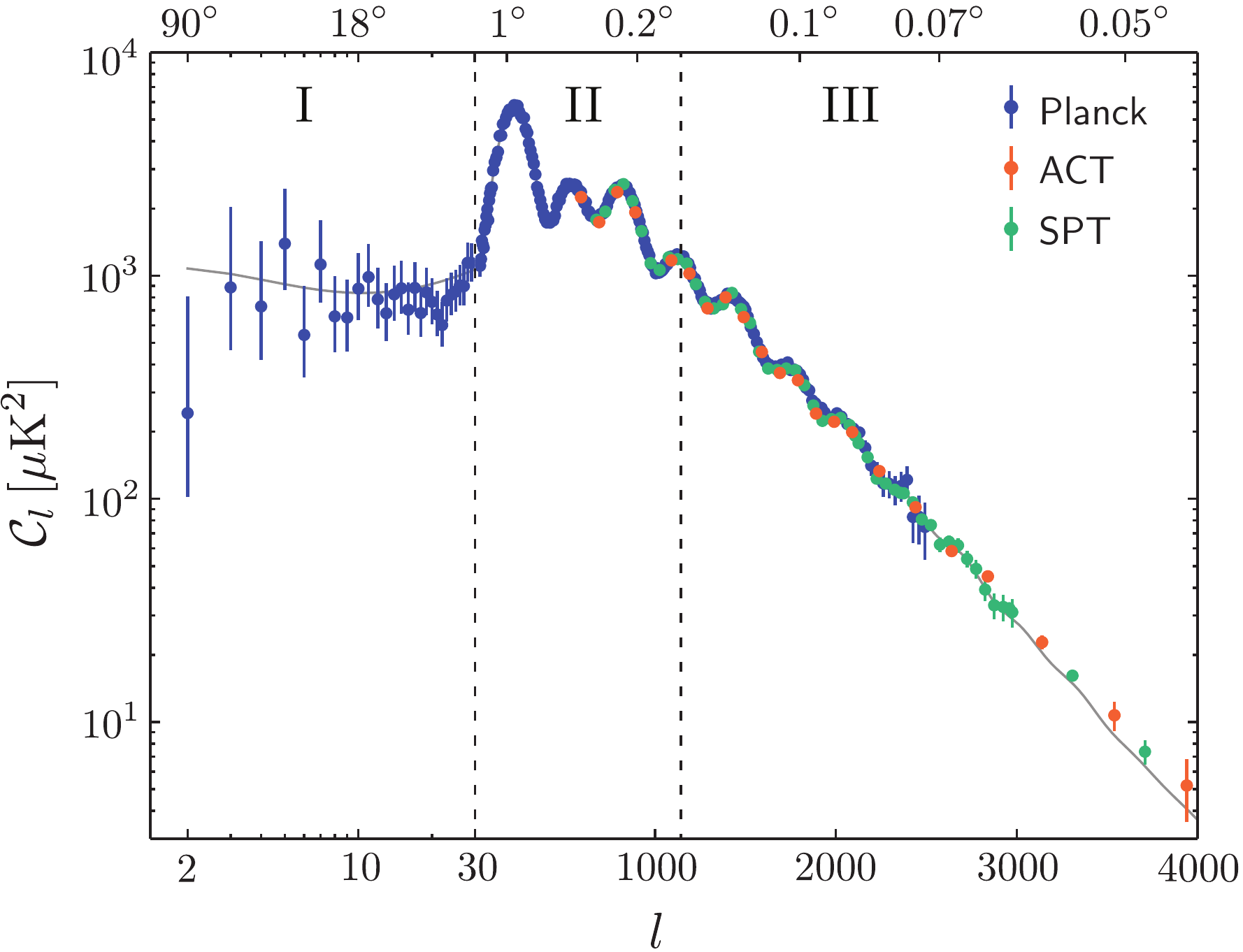} 
    \caption{The CMB spectrum ${\cal C}_l \equiv l(l + 1)C_l$ as
      observed by Planck~\cite{Aghanim:2018eyx}, the South Pole Telescope (SPT)~\cite{Calabrese:2013jyk}, and the Atacama
      Cosmology Telescope (ACT)~\cite{Louis:2016ahn}. The angular variations of the CMB power spectrum are
    consequence of the dynamics of sound waves in the photon-baryon
    fluid. On large scales (region I), the fluctuations are frozen and
    we directly see the spectrum of the initial conditions. At
    intermediate scales (region II), we observe the oscillations of
    the fluid as captured at the moment of last-scattering. Finally,
    on small scales (region III), fluctuations are damped because
    their wavelengths are smaller than the mean free path of the
    photons. This figure is courtesy of Daniel Baumann and has been
    published in~\cite{Baumann:2018muz}.}
\label{fig:dos}
\end{figure}

As a matter of choice, we can also extrapolate the value of $H_0$ from
cosmological observations, particularly from measurements of
temperature and polarization anisotropies in the
CMB~\cite{Jungman:1995bz}. These anisotropies encode information on
the relativistic energy density at the surface of last-scattering. In CMBology $\theta_*$ is the angular size of the sound horizon at
recombination. This angular scale can be
inferred from the anisotropy power spectrum shown in Fig.~\ref{fig:dos}. Deducing the $H_0$
value from $\theta_*$ requires a model to describe the
expansion history of the universe both before the radiation decouples
from matter and since that decoupling. For $\Lambda$CDM, the
Planck Collaboration finds $H_0 = 67.4 \pm 0.5~{\rm km} \, {\rm s}^{-1} \, {\rm
  Mpc}^{-1}$~\cite{Aghanim:2018eyx}.

Adding to the story, the CMB measurements can be combined with probes of the expansion
history at lower redshift, such as baryon acoustic oscillation (BAO)
or SNe Ia distance measurements for a cross calibration of the cosmic
distance ladder~\cite{Cuesta:2014asa}. While standard candles,
calibrated from the local measurement of $H_0$ provide a ``direct''
cosmic distance ladder (from nearby out towards cosmological
distances), the BAO provides an ``inverse'' cosmic distance ladder,
calibrated using the inferred $\Lambda$CDM sound-horizon scale at the
surface of last scattering and extended in, towards lower
redshifts. SN Ia and BAO measurements overlap in redshift and so the
direct and inverse cosmic distance ladders can be calibrated off one
another. The statistical errors in both distance measures as function
of redshifts are reaching percent level.

The latest chapter in the story is courtesy of the Carnegie-Chicago Hubble Program. As we can see in Fig.~\ref{fig:1} the latest $H_0$ measurement, with a
calibration of SNe Ia that is based on the tip of the red-giant branch
distance scale, falls in between the previous determinations:
$H_0 = 69.8 \pm 1.9~{\rm km} \, {\rm s}^{-1} \, {\rm Mpc}^{-1}$ is
between $1.2\sigma$ of the CMB result and it is also consistent at
better than $2\sigma$ with the Cepheid distance scale
measurement~\cite{Freedman:2020dne}. Independent measurements using
time-delays of multiply imaged quasars by the H0LiCOW Collaboration
and gravitational waves by the LIGO-Virgo Collaboration lead to
$H_0 = 73.3^{+1.7}_{-1.8}~{\rm km} \, {\rm s}^{-1} \, {\rm
  Mpc}^{-1}$~\cite{Wong:2019kwg} and
$H_0 = 68^{+14}_{-7}~{\rm km} \, {\rm s}^{-1} \, {\rm
  Mpc}^{-1}$~\cite{Abbott:2019yzh}, respectively. Remarkably, the
study of statistically independent datasets shows that the
significance of the discrepancy between local $H_0$ measurements and
the early universe prediction is $4.4\sigma$~\cite{Verde:2019ivm},
providing strong evidence for physics beyond $\Lambda$CDM.\\

{\bf \color{rossoCP3} \underline{THE HUNT FOR LIGHT RELICS:}} The $SU(3)_C
\otimes SU(2)_L \otimes U(1)_Y$ standard model (SM) of
particle physics has recently endured intensive scrutiny, with a
dataset corresponding to a total integrated luminosity of
$66~{\rm fb}^{-1}$ of $pp$ collisions at $\sqrt{s} = 13~{\rm TeV}$,
and it has proven once again to be a remarkable structure that is
consistent with all experimental results by tuning more or less 19
free parameters~\cite{Tanabashi:2018oca}. However, the SM is inherently an incomplete theory,
as it does not explain all known fundamental physical
phenomena: the most obvious omission is that it does not provide a
unification with gravity. Knowing that the SM is incomplete leads us to search for new
fundamental particles. On the one hand, the search for heavy particles will continue at the LHC in 2021. On the other hand, we have seen that
CMB anisotropies are sensitive to the relativistic energy density at
recombination, and so from this we can gain information about the
number of light species at the surface of last-scattering.

The presence of any additional light species with $g$
degrees of freedom is usually characterized by
\begin{eqnarray}
\Delta N_{\rm eff}  & \equiv &  N_{\rm eff} -
N_{\rm eff}^{\rm SM} \nonumber \\ & =  &  g \ \left(\frac{10.75}{g_*
    (T_{\rm dec})}\right)^{4/3} \times \left\{ \begin{array}{cl} 4/7
                                                &~~{\rm boson} \\
                                                1/2&~~{\rm
                                                     fermion} \end{array}
                                                 \right. \,,
\label{eq:Neff}
\end{eqnarray}
where
\begin{equation}
N_{\rm eff} \equiv \frac{\rho_{\rm R} -
\rho_\gamma}{\rho_{\nu_L}} 
\label{neff}
\end{equation}
is the number of ``equivalent'' light
neutrino species in units
of the density of a single Weyl neutrino
\begin{equation}
\rho_{\nu_L} =\frac{7 \pi^2}{120} \
\left(\frac{4}{11}\right)^{4/3} \ T_\gamma^4 \,,
\end{equation}
$\rho_\gamma$ is the energy density of
photons (with temperature $T_\gamma$), $\rho_{\rm R}$ is the total
energy density in relativistic particles, $T_{\rm dec}$ is the temperature at which
particle species decouple from the primordial plasma,  and the function
$g_*(T_{\rm dec})$ is the number of effective degrees of freedom
(defined as the number of independent states with an additional factor
of 7/8 for fermions) of the SM particle content at the temperature
$T_{\rm dec}$~\cite{Steigman:1977kc}.\footnote{If relativistic particles are present that have decoupled from the photons, it is necessary to distinguish between two kinds of $g_*$: $g_\rho$ which is associated with the total energy density, and $g_s$ which is associated with the total entropy density~\cite{Kolb:1990vq}. For our calculations we use $g_* = g_\rho = g_s$.} The
normalization of $N_{\rm eff}$ is such that it gives $N_{\rm eff}^{\rm
  SM} = 3.046$ for three families of massless
$\nu_L$~\cite{Mangano:2005cc}. Note that the SM value slightly exceeds the
integer 3 mainly because neutrinos do not decouple instantaneously,
and this enables them to share some of the energy released by  $e^+ e^-$ annihilations.

\begin{table}
\caption{Effective numbers of degrees of freedom in the SM. $T_{\rm crit}$ 
indicates the critical temperature of the confinement-deconfinement transition between
 quarks and hadrons~\cite{Kolb:1990vq}.}
\begin{tabular}{llc}
  \hline
  \hline
{Temperature} & { New particles} \qquad
&$4g_*(T)$ \\
\hline\rule{0pt}{12pt}
$T < m_{ e}$   &     $\gamma$ +   $\nu_{e,\mu,\tau} \overline
                 \nu_{e, \mu, \tau}$ & 29 \\
$m_{ e} <   T  < m_\mu$ &    $e^{\pm}$ & 43 \\
$m_\mu <  T  < m_\pi$  &   $\mu {}^{\pm}$ & 57 \\
$m_\pi <  T < T_{\rm crit}$  & $\pi^0, \pi^\pm$ & 69 \\
$T_{\rm crit} <  T  < m_c$~~~~~~~~~~&
   $- \pi^0,\pi^\pm$ + $  u,{\bar u},d,{\bar d},s,{\bar s}$, $g$~~~~~~~~~~ &  ~~~~~~247~~~~~~ \\
$m_{ c} <  T < m_\tau$ &  $c,{\bar c}$ & 289 \\
$m_\tau < T < m_{b}$ & $\tau {}^{\pm}$ & 303 \\
$m_{ b} < T < m_{ W,Z}$ & $b,{\bar b}$ & 345 \\
$m_{ W,Z} <  T < m_{H}$ & $W^{\pm}, Z$ & 381 \\
$ m_H< T < m_{t}$ & $H^0$ & 385 \\
$m_t< T $ & $t,{\bar t}$  & 427 \\
\hline \hline
\end{tabular}
 \label{tabla:gT}
\end{table}

The change in $g_*(T)$ (ignoring mass effects) is given in
Table~\ref{tabla:gT}. Comparing the 106.75 degrees of freedom of the
SM with the 10.75 degrees of freedom of the primordial plasma before
neutrino decoupling it is straightforward to see that for a massless
(real) spin-$0$ scalar, spin-$\tfrac{1}{2}$ (Weyl) fermion, and
massive spin-$1$ vector boson the contributions to $N_{\rm eff}$
asymptote to specific values of $\Delta N_{\rm eff} = 0.027$, $0.047$,
and $0.080$; respectively. (Asymptote here refers to relativistic
species decoupling just before $t \bar t$ freeze-out.)

\begin{figure}[tb]
\centering
\includegraphics[width=0.97\columnwidth]{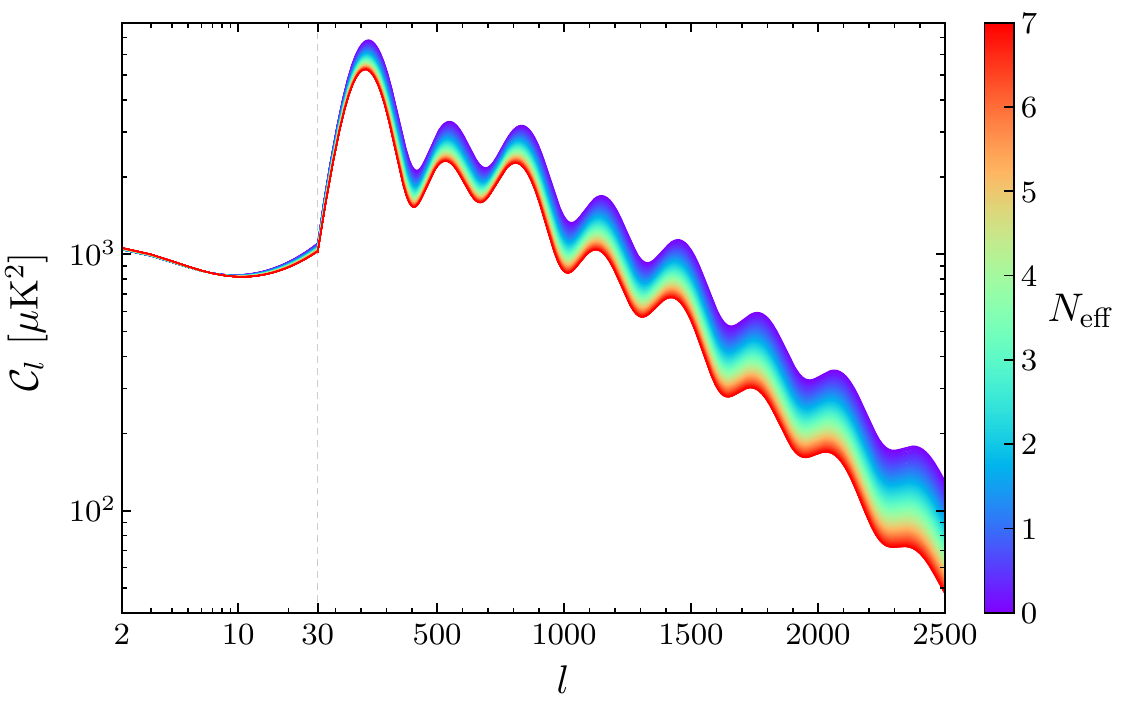}
    \caption{Variation of the CMB spectrum ${\cal C}_l \equiv l(l +
      1)C_l$ as a function of $N_{\rm eff}$ for fixed $\theta_*$. This figure is courtesy of Daniel Baumann and has been
      published in~\cite{Baumann:2018muz}.}
    \label{fig:tres}
\end{figure}

\begin{figure*}[tb] 
\centering
\includegraphics[width=1.97\columnwidth]{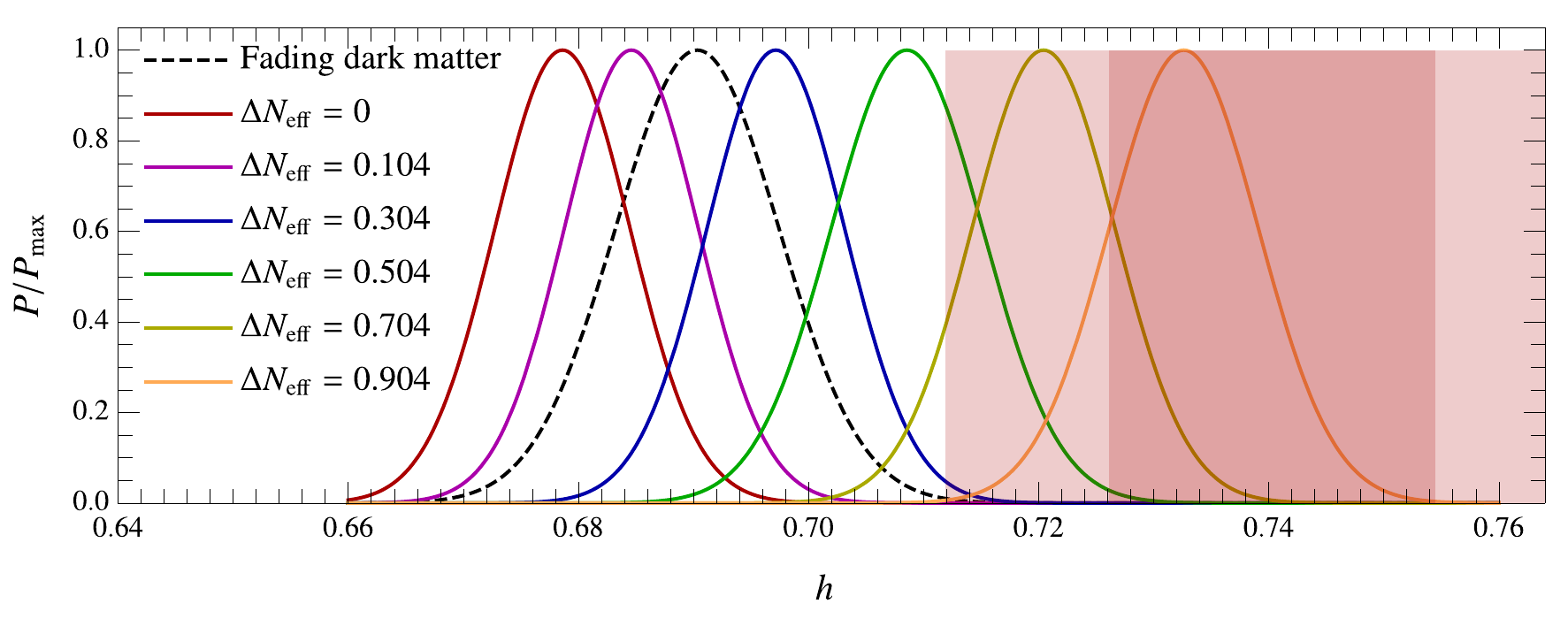}
    \caption{Rescaled posterior distributions of $h$ (due to marginalization over additional free parameters) with different
      choices of $N_{\rm eff}$ in the $\Lambda$CDM  6 parameter fit 
      of~\cite{Vagnozzi:2019ezj}. The rescaled posterior distribution
      of $h$ for the best fit ($\tilde c = 0.3$) of the fading dark
      matter model is indicated with the dashed curve~\cite{Agrawal:2019dlm}.The
      shaded areas indicate the $1\sigma$ and $2\sigma$ regions as
      determined by SH0ES~\cite{Riess:2019cxk}.
\label{fig:cuatro}}
\end{figure*}

As an aftermath, any contribution of light relics to the energy
density leads to observable consequences in the CMB temperature and
polarization anisotropy. As shown in Fig.~\ref{fig:tres},  the main
effect of $\Delta N_{\rm eff} >0$ is to increase the damping of the
CMB spectrum. Increasing $N_{\rm eff}$ also increases $H(z_*)$, the
expansion rate at recombination. The main limiting factor in
constraining $\Delta N_{\rm eff}$ from CMB measurements of $\theta_*$
is a degeneracy with the primordial helium fraction $Y_P \equiv n_{\rm
  He}/n_b$. For a fixed fractional density of baryons $\Omega_bh^2$, increasing $Y_P$ reduces the power
in the damping tail. In other words, the parameters $Y_P$ and $N_{\rm eff}$ are anti-correlated~\cite{Baumann:2018muz}. Combining  CMB, BAO,
and BBN observations and considering a single-parameter extension to
the based-$\Lambda$CDM model the Planck Collaboration reported $N_{\rm eff} =
3.04^{+0.22}_{-0.22}$, which translates into a limit of $\Delta N_{\rm
  eff} < 0.214$ at the 95\%~CL~\cite{Aghanim:2018eyx}.  This limit combines the helium measurements of~\cite{Aver:2015iza,Peimbert:2016bdg} with the latest
  deuterium abundance measurements of~\cite{Cooke:2017cwo} using the
  the \texttt{PArthENoPE} code~\cite{Pisanti:2007hk} 
considering $d(p,\gamma)^3{\rm He}$ reaction rates
from~\cite{Marcucci:2015yla}. Should they instead
use the helium abundance measurement of~\cite{Izotov:2014fga} in place
of~\cite{Aver:2015iza,Peimbert:2016bdg} would have lead to $N_{\rm eff} = 3.37 \pm
0.22$~\cite{Aghanim:2018eyx}. This gives 
 a 95\% CL limit of 
 $\Delta N_{\rm eff} < 0.544$, which is in $2.9 \sigma$ tension with $N_{\rm
   eff}^{\rm SM}$. Both these bounds have the power to
 exclude many beyond SM physics models (e.g.~\cite{Anchordoqui:2019yzc}).

 Now, we can ask ourselves whether  $\Delta N_{\rm eff} > 0$ which increases $H(z_*)$ can
 solve the trouble with $H_0$. In Fig.~\ref{fig:cuatro} we show the
 normalized posterior distributions of $h$ for different choices of
 $N_{\rm eff}$ in the $\Lambda$CDM 6 parameter fit of~\cite{Vagnozzi:2019ezj}. It
 is evident that the most restrictive 95\% CL upper limit on
 $\Delta N_{\rm eff}$ from the combination of CMB, BAO, and BBN
 observations~\cite{Aghanim:2018eyx} severely constrains a solution of
 the $H_0$ problem in terms of additional relativistic degrees of
 freedom. Consideration of the larger helium abundance measured
 in~\cite{Izotov:2014fga}, still precludes a full solution of the
 $H_0$ problem in terms of additional light species at the CMB epoch.\\

{\bf \color{rossoCP3} \underline{ECHOES OF VIBRATING STRINGS?}} String theory is the most promising candidate for a consistent quantum
theory of gravitationally interacting matter fields. Therefore, it of
interest to explore whether stringy models can solve the $H_0$
problem. Realizing de Sitter vacua in string theory is challenging. A
varying dark energy seems more string theory friendly. This is because
string compactifications allow for quintessence fields $\phi$ that can
be the source of dark energy, in the sense that the field's potential
$V(\phi)$ is equal to the density of dark energy $\Lambda$. String
theory calculations, however, have suggested that the slope of this
potential $V'$ must be nonzero~\cite{Obied:2018sgi}, calling $\Lambda$CDM into
question~\cite{Agrawal:2018own}. All the same, quintessence models
exacerbate the $H_0$ tension~\cite{Raveri:2018ddi,Colgain:2019joh}. A
possible bolt-hole to this conclusion is to consider a (non-trivial) coupling between dark matter and dark energy,
 \begin{equation}
   m(\phi) \propto \exp \{- \tilde c \, \phi\} \,,                          \end{equation}
 where $\tilde c \sim {\cal O} (1)$ in Planck units~\cite{Agrawal:2019dlm}.  Such a
 coupling leads to  fading of dark matter in the recent cosmological epoch which is compensated by a
 bigger value of dark energy. A definite realization of the cosmological string framework of fading
 dark matter has been given elsewhere~\cite{Anchordoqui:2019amx}. As shown in Fig.~\ref{fig:cuatro} the fading dark matter hypothesis automatically
 produces larger values of $H_0$ than $\Lambda$CDM, relieving tensions
 in the data but not fully resolving them. Indeed the best fit value ($\tilde c = 0.3$) yields $H_0 = 69.06^{+ 0.66}_{-0.73}~{\rm km} \, {\rm s}^{-1} \, {\rm
  Mpc}^{-1}$~\cite{Agrawal:2019dlm}, which is characteristic of all models with late dark energy
modification of the $\Lambda$CDM expansion
history~\cite{Salvatelli:2014zta,Kumar:2017dnp,DiValentino:2017iww,Yang:2018euj,Kumar:2019wfs,DiValentino:2019ffd,DiValentino:2019jae}.
 This is because
the local distance ladder calibrates SNe Ia far into the Hubble flow and
if dark matter fading takes place too recently then it would raise
$H_0$ but without actually changing the part of the Hubble diagram
where the tension is inferred. More concretely, by substituting the
SH0ES calibration to the Pantheon SNe Ia dataset, the ability of late times dark energy transitions  to reduce the Hubble tension drops
effectively to $H_0 = 69.17 \pm 1.09~{\rm km} \, {\rm s}^{-1} \, {\rm
  Mpc}^{-1}$~\cite{Benevento:2020fev}.

Dirac neutrino masses, which are ubiquitous in
 intersecting D-brane string
 compactifications~\cite{Blumenhagen:2005mu,Blumenhagen:2006ci}, might
 come to the rescue~\cite{Anchordoqui:2011nh}. After looking over the
 distributions shown in Fig.~\ref{fig:cuatro} we can argue that the combined
 effect produced by fading dark
 matter and extra
 effective number of neutrino generations at the CMB appears to have the potential to accommodate the $H_0$ tension if  $\Delta N_{\rm
   eff} \sim 0.5$.  Moreover,
 intersecting D-brane models typically include  enlarged gauge
 sectors which are broken down to the SM gauge symmetry. If the
 symmetry breaking scale is not too high, the associated heavy
 $Z'$ gauge bosons could be within the LHC
 reach, thereby relating the Hubble tension to precision measurements at
 colliders. From (\ref{eq:Neff}) and Table~\ref{tabla:gT}  it is
 straightforward to see that for Dirac neutrino masses, the 95\% CL upper bound $\Delta N_{\rm eff} <
 0.214$ prevents $\nu_R$ decoupling at temperatures below the $b\bar
 b$ freeze-out. If we instead consider the less restrictive upper bound  $\Delta N_{\rm eff} < 0.544$ the
 three right-handed neutrinos could decouple at  the QCD crossover
 transition, such that  the mass scale of the associated $Z'$ gauge
 boson could be probed by the LHC
 Run3.\\

{\bf \color{rossoCP3} \underline{THE DRAMATIS PERSONAE:}} We end with an update of the results given in~\cite{Anchordoqui:2011nh,Anchordoqui:2012wt,Anchordoqui:2012qu}. To develop our program in the simplest way, we will work within the
construct of a minimal model. The gauge-extended $U(1)_C \otimes Sp(1)_L \otimes U(1)_{I_R} \otimes U(1)_L$ D-brane model has the attractive property of
elevating the two major global symmetries of the SM (baryon number $B$ and lepton
number $L$) to local gauge symmetries~\cite{Cremades:2003qj,Anchordoqui:2011eg}. The $U(1)_L$ symmetry prevents
the generation of Majorana masses, leading to three superweakly interacting right-handed neutrinos. This also renders a $B - L$
symmetry non-anomalous. We now use the upper limit on $\Delta N_{\rm
  eff}$ derived by the Planck Collaboration to show that the superweak interactions of these Dirac states (through
their coupling to the TeV-scale $I_R$ gauge boson) permit
right-handed neutrino decoupling on the QCD crossover transition.

\begin{figure*}[!t]
\centering
\includegraphics[width=1.97\columnwidth]{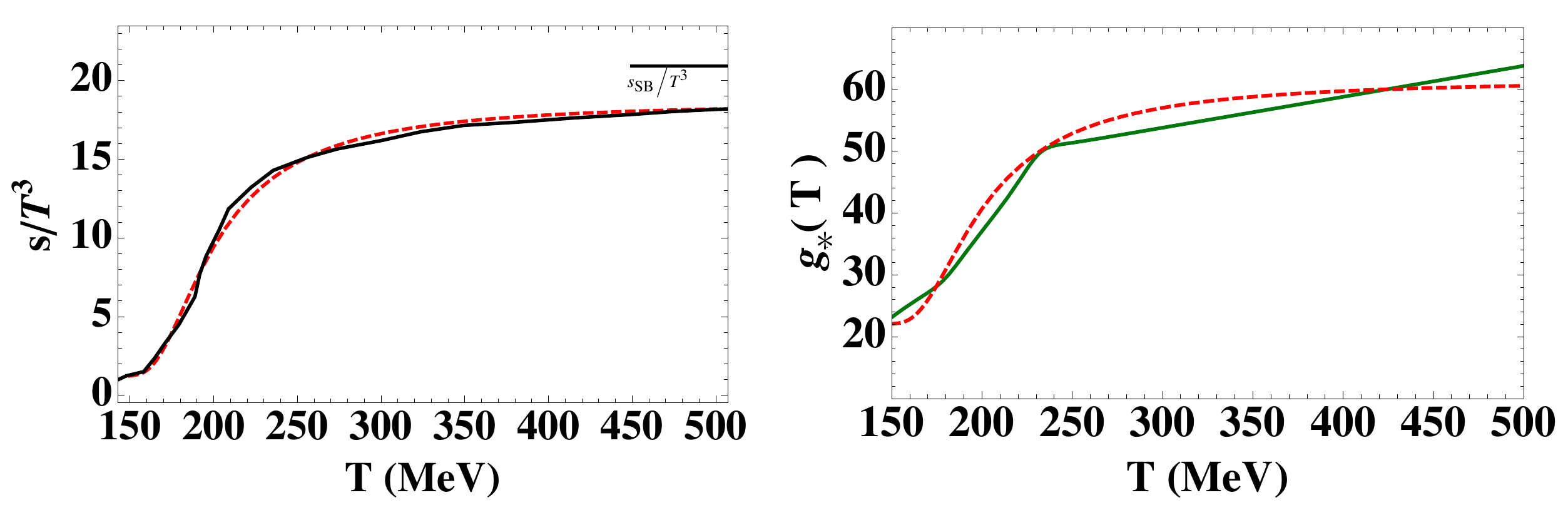}
  \caption{{\bf Left.} The parametrization of the entropy density
    given in~(\ref{soverT})  (dashed line) superposed on the
    result from high statistics lattice
    simulations~\cite{Bazavov:2009zn} (solid line). {\bf Right.}~Comparison of $g_* (T)$ obtained using~(\ref{gdet}) (dashed
    line) and the phenomenological estimate of~\cite{Laine:2006cp,Steigman:2012nb} (solid line).}
\label{fig:cinco}
\end{figure*}

 As indicated in Table~\ref{tabla:gT} at energies  above the deconfinement transition towards the quark gluon
plasma, quarks and gluons are the relevant fields for the QCD sector,
such that the total number of SM relativistic degrees of freedom is $g_* = 61.75$.  As the
universe cools down, the SM plasma transitions to a regime where
mesons and baryons are the pertinent degrees of freedom. Precisely,
the relevant hadrons present in this energy regime are pions and
charged kaons, such that $g_* = 19.25$~\cite{Brust:2013xpv}. This
significant reduction in the degrees of freedom results from the rapid
annihilation or decay of any more massive hadrons which may have
formed during the transition. The quark-hadron crossover transition
therefore corresponds to a large redistribution of entropy into the
remaining degrees of freedom. Concretely, the effective number of
interacting relativistic degrees of freedom in the plasma at temperature $T$ is given by 
\begin{equation}
g_*
(T) \simeq r (T) \left(g_B+ \frac{7}{8} g_F \right),
\end{equation}
with $g_B = 2$ for each real vector field and $g_F = 2$
for each spin-$\frac{1}{2}$ Weyl field~\cite{Anchordoqui:2014dpa}. The coefficient $r (T)$ is
unity for leptons, two for photon contributions, and is the ratio $s(T
)/s_{\rm SB}$ for the quark-gluon plasma. Here, $s(T)$ $(s_{\rm SB})$
is the actual (ideal Stefan-Boltzmann) entropy shown in
Fig~\ref{fig:dos}. For $150~{\rm MeV} < T < 500~{\rm MeV}$,  we
parametrize the entropy rise during the confinement-deconfinement
changeover by
\begin{eqnarray}
\frac{s}{T^3} & \simeq & \frac{42.82}{\sqrt{392 \pi}} \exp \left[-\frac{\left(
      T_{\rm MeV} - 151 \right)^2}{392} \right]  + \left( \frac{195.1}{T_{\rm
      MeV} - 134} \right)^2  \nonumber \\
& \times & 18.62 \ \frac{\exp[195.1/(T_{\rm MeV} -134)]} {\left\{
    \exp[195.1/(T_{\rm MeV} -134)] - 1\right\}^2} \, .
\label{soverT}
\end{eqnarray}
For the same energy range, we obtain
\begin{equation}
g_*(T) \simeq 47.5 \ r(T) + 19.25 \, .
\label{gdet}
\end{equation}
In Fig.~\ref{fig:dos} we show $g_*(T)$ as given by (\ref{gdet}). Our
parametrization is in very good agreement with the phenomenological
estimate of~\cite{Laine:2006cp,Steigman:2012nb}. Using (\ref{eq:Neff}) and (\ref{gdet}), and considering 3 $\nu_R$ flavors we
determine the $T_{\rm dec}$ from the plasma for the ``observed'' central value $\Delta N_{\rm eff}
=0.324$ and the 95\% CL upper limit $\Delta N_{\rm eff} = 0.544$,
which bracket the required $\Delta N_{\rm eff} \sim 0.5$.

\begin{figure}[tb]
\centering
\includegraphics[width=0.97\columnwidth]{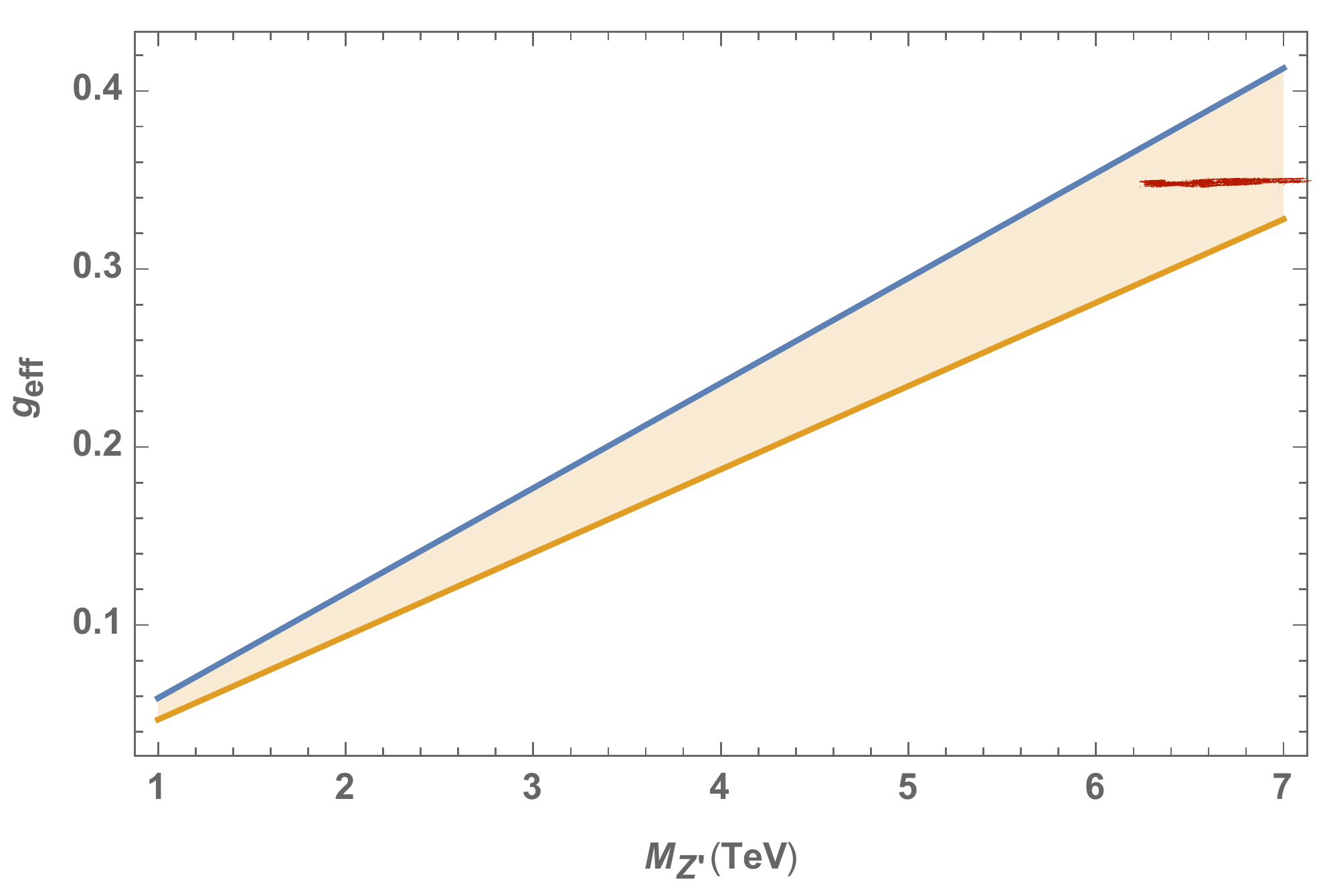}
  \caption{The shaded region brackets the function $g_{\rm eff}
    (M_{Z'})$ in the interval  $0.324< \Delta N_{\rm
  eff} < 0.544$. The condition of decoupling for thermal equilibrium
has been imposed. The horizontal line indicates
the effective coupling of a D-brane model in which $Z'$ couples mostly
to the third component of a right-handed isospin $I_R$. The chiral
couplings of this gauge boson are tabulated in~\cite{Anchordoqui:2012wt}. Termination of the horizontal line on the left reflects the LHC experimental limit on the mass of the gauge boson~\cite{CMS:2019tbu,Aad:2019hjw,Aad:2020kep}.} 
\label{fig:seis}
\end{figure}

We now turn to use $T_{\rm dec}$ in conjunction with the decoupling
condition to constrain the masses and couplings of the heavy gauge
bosons. The physics of interest  
will be taking place at energies in the region of the quark-hadron
crossover transition, so that we will restrict ourselves to the
following fermionic fields, and their contribution to relativistic
degrees of freedom: $[3u_R] + [3d_R] + [3s_R] + [3\nu_L +e_L +\mu_L] +
[e_R +\mu_R] + [3u_L + 3d_L + 3s_L] + [3\nu_R]$. This amounts to 28
Weyl fields, translating to 56 fermionic relativistic degrees of
freedom. The right-handed neutrino decouples from the plasma when its
mean free path becomes greater than the Hubble radius at that time. We
calculate the $\nu_R$
interaction rate $\Gamma(T)$ and via the prescription $\Gamma (T_{\rm
  dec} ) = H (T_{\rm dec})$ we obtain the desired constraint,
\begin{equation}
  \frac{g_{\rm eff}}{M_{Z'}} = \left(\frac{3}{\Delta N_{\rm eff}}
  \right)^{3/32} \left(\frac{17.41}{M_{\rm Pl} \     T_{\rm dec}^3} \right)^{1/4} \,,
\end{equation}  
where $g_{\rm eff}$ is the effective coupling of the $Z'$ gauge boson,
$M_{Z'}$ its mass, and $M_{\rm Pl}$ is the Planck
mass~\cite{Anchordoqui:2011nh,Anchordoqui:2012wt,Anchordoqui:2012qu}. In
Fig.~\ref{fig:seis} we show the $g_{\rm
  eff}-M_{Z'}$ parameter space consistent with $0.324< \Delta N_{\rm
  eff} < 0.544$. The model is fully predictive and can be confronted
with future data from LHC Run3.\\

{\bf \color{rossoCP3} \underline{THE TAKE HOME MESSAGE:}} Solving the Hubble ($\Lambda$CDM) tension is very much an ongoing enterprise. The resolution of this conundrum will likely require
a coordinated effort from the side of theory, interpretation, data
analysis, and observation: CMBPol is expected to reach a $2\sigma$
precision of $\Delta N_{\rm eff} = 0.09$~\cite{Galli:2010it} and CMB-S4 is expected to reach a $2\sigma$ precision of
$\Delta N_{\rm eff} = 0.06$~\cite{Abazajian:2019eic}. If the past is
any guidepost to the future, we can expect surprising results connecting string theory to
data. 

\section*{Acknowledgments} Thanks are due to Daniel Baumann for
permission to reproduce Figs.~\ref{fig:dos} and \ref{fig:tres}. This work has been
    supported by the U.S. National Science Foundation (NSF
    Grant PHY-1620661) and the National Aeronautics and Space
    Administration (NASA Grant 80NSSC18K0464).  This write-up is based
    on a talk given at the Phenomenology 2020 Symposium, University of
    Pittsburgh, 4-6 May 2020. Any opinions,
    findings, and conclusions or recommendations expressed in this
    material are those of the author and do not necessarily reflect
    the views of the NSF or NASA.


\begin{thebibliography}{99}

\bibitem{Freedman:2000cf} 
  W.~L.~Freedman {\it et al.} [HST Collaboration],
    {\color{rossoCP3} Final results from the Hubble Space Telescope key project to measure the Hubble constant},
  Astrophys.\ J.\  {\bf 553}, 47 (2001)
  doi:10.1086/320638
  [astro-ph/0012376].


\bibitem{Riess:2009pu} 
  A.~G.~Riess {\it et al.},
    {\color{rossoCP3} A redetermination of the Hubble constant with the Hubble Space Telescope from a differential distance ladder},
  Astrophys.\ J.\  {\bf 699}, 539 (2009)
  doi:10.1088/0004-637X/699/1/539
  [arXiv:0905.0695 [astro-ph.CO]].

  
 
\bibitem{Riess:2011yx} 
  A.~G.~Riess {\it et al.},
    {\color{rossoCP3} A 3\% solution: Determination of the Hubble constant with the Hubble Space Telescope and Wide Field Camera 3},
  Astrophys.\ J.\  {\bf 730}, 119 (2011)
  Erratum: [Astrophys.\ J.\  {\bf 732}, 129 (2011)]
  doi:10.1088/0004-637X/732/2/129, 10.1088/0004-637X/730/2/119
  [arXiv:1103.2976 [astro-ph.CO]].


\bibitem{Freedman:2012ny} 
  W.~L.~Freedman, B.~F.~Madore, V.~Scowcroft, C.~Burns, A.~Monson, S.~E.~Persson, M.~Seibert and J.~Rigby,
    {\color{rossoCP3} Carnegie Hubble Program: A mid-infrared calibration of the Hubble Constant},
  Astrophys.\ J.\  {\bf 758}, 24 (2012)
  doi:10.1088/0004-637X/758/1/24
  [arXiv:1208.3281 [astro-ph.CO]].

  
  
\bibitem{Riess:2016jrr} 
  A.~G.~Riess {\it et al.},
    {\color{rossoCP3} A 2.4\% determination of the local value of the Hubble constant},
  Astrophys.\ J.\  {\bf 826}, no. 1, 56 (2016)
  doi:10.3847/0004-637X/826/1/56
  [arXiv:1604.01424 [astro-ph.CO]].


\bibitem{Riess:2018byc} 
  A.~G.~Riess {\it et al.},
    {\color{rossoCP3} Milky Way Cepheid standards for measuring cosmic distances and application to Gaia DR2: Implications for the Hubble constant},
  Astrophys.\ J.\  {\bf 861}, no. 2, 126 (2018)
  doi:10.3847/1538-4357/aac82e
  [arXiv:1804.10655 [astro-ph.CO]].

 
\bibitem{Riess:2019cxk} 
  A.~G.~Riess, S.~Casertano, W.~Yuan, L.~M.~Macri and D.~Scolnic,
    {\color{rossoCP3} Large Magellanic Cloud Cepheid standards provide a 1\% foundation for the determination of the Hubble constant and stronger evidence for physics beyond $\Lambda$CDM},
  Astrophys.\ J.\  {\bf 876}, no. 1, 85 (2019)
  doi:10.3847/1538-4357/ab1422
  [arXiv:1903.07603 [astro-ph.CO]].

  
\bibitem{Spergel:2003cb} 
  D.~N.~Spergel {\it et al.} [WMAP Collaboration],
    {\color{rossoCP3} First year Wilkinson Microwave Anisotropy Probe (WMAP) observations: Determination of cosmological parameters},
  Astrophys.\ J.\ Suppl.\  {\bf 148}, 175 (2003)
  doi:10.1086/377226
  [astro-ph/0302209].



\bibitem{Spergel:2006hy} 
  D.~N.~Spergel {\it et al.} [WMAP Collaboration],
    {\color{rossoCP3} Wilkinson Microwave Anisotropy Probe (WMAP) three year results: Implications for cosmology},
  Astrophys.\ J.\ Suppl.\  {\bf 170}, 377 (2007)
  doi:10.1086/513700
  [astro-ph/0603449].


\bibitem{Komatsu:2008hk} 
  E.~Komatsu {\it et al.} [WMAP Collaboration],
    {\color{rossoCP3} Five-year Wilkinson Microwave Anisotropy Probe (WMAP) observations: Cosmological interpretation},
  Astrophys.\ J.\ Suppl.\  {\bf 180}, 330 (2009)
  doi:10.1088/0067-0049/180/2/330
  [arXiv:0803.0547 [astro-ph]].

 
\bibitem{Komatsu:2010fb}
  E.~Komatsu {\it et al.} [WMAP Collaboration],
    {\color{rossoCP3} Seven-year Wilkinson Microwave Anisotropy Probe (WMAP) observations: Cosmological interpretation},
  Astrophys.\ J.\ Suppl.\  {\bf 192}, 18 (2011)
  doi:10.1088/0067-0049/192/2/18
  [arXiv:1001.4538 [astro-ph.CO]].


 \bibitem{Hinshaw:2012aka} 
  G.~Hinshaw {\it et al.} [WMAP Collaboration],
    {\color{rossoCP3} Nine-year Wilkinson Microwave Anisotropy Probe (WMAP) observations: Cosmological parameter results},
  Astrophys.\ J.\ Suppl.\  {\bf 208}, 19 (2013)
  doi:10.1088/0067-0049/208/2/19
  [arXiv:1212.5226 [astro-ph.CO]].


\bibitem{Ade:2013zuv} 
  P.~A.~R.~Ade {\it et al.} [Planck Collaboration],
    {\color{rossoCP3} Planck 2013 results XVI: Cosmological parameters},
  Astron.\ Astrophys.\  {\bf 571}, A16 (2014)
  doi:10.1051/0004-6361/201321591
  [arXiv:1303.5076 [astro-ph.CO]].


\bibitem{Ade:2015xua} 
  P.~A.~R.~Ade {\it et al.} [Planck Collaboration],
    {\color{rossoCP3} Planck 2015 results XIII: Cosmological parameters},
  Astron.\ Astrophys.\  {\bf 594}, A13 (2016)
  doi:10.1051/0004-6361/201525830
  [arXiv:1502.01589 [astro-ph.CO]].


\bibitem{Abbott:2017smn} 
  T.~M.~C.~Abbott {\it et al.} [DES Collaboration],
    {\color{rossoCP3} Dark Energy Survey year 1 results: A precise $H_0$ estimate from DES Y1, BAO, and D/H data},
  Mon.\ Not.\ Roy.\ Astron.\ Soc.\  {\bf 480}, no. 3, 3879 (2018)
  doi:10.1093/mnras/sty1939
  [arXiv:1711.00403 [astro-ph.CO]].


\bibitem{Aghanim:2018eyx} 
  N.~Aghanim {\it et al.} [Planck Collaboration],
    {\color{rossoCP3} Planck 2018 results VI: Cosmological parameters}
  arXiv:1807.06209 [astro-ph.CO].
  
\bibitem{Jang:2015} 
  I.~S.~Jang and M.~G.~Lee,
  {\color{rossoCP3} The tip of the red giant branch distances to type Ia supernova host galaxies III: NGC 4038/39 and NGC 5584},
 Astrophys.\ J.\  {\bf 807}, no. 2, 133 (2015)
doi:10.1088/0004-637X/807/2/133
[arXiv:1506.03089 [astro-ph.GA]].
  


\bibitem{Jang:2017dxn} 
  I.~S.~Jang and M.~G.~Lee,
    {\color{rossoCP3} The tip of the red giant branch distances to type Ia supernova host galaxies V: NGC 3021, NGC 3370, and NGC 1309 and the value of the Hubble constant},
  Astrophys.\ J.\  {\bf 836}, no. 1, 74 (2017)
  doi:10.3847/1538-4357/836/1/74
  [arXiv:1702.01118 [astro-ph.CO]].


\bibitem{Freedman:2019jwv} 
  W.~L.~Freedman {\it et al.},
    {\color{rossoCP3} The Carnegie-Chicago Hubble Program VIII: An independent determination of the Hubble constant based on the tip of the red giant branch},
  doi:10.3847/1538-4357/ab2f73
  arXiv:1907.05922 [astro-ph.CO].



\bibitem{Freedman:2020dne} 
  W.~L.~Freedman {\it et al.},
    {\color{rossoCP3} Calibration of the tip of the red giant branch (TRGB)},
  doi:10.3847/1538-4357/ab7339
  arXiv:2002.01550 [astro-ph.GA].



\bibitem{Jungman:1995bz} 
  G.~Jungman, M.~Kamionkowski, A.~Kosowsky and D.~N.~Spergel,
  {\color{rossoCP3} Cosmological parameter determination with microwave background maps},
  Phys.\ Rev.\ D {\bf 54}, 1332 (1996)
  doi:10.1103/PhysRevD.54.1332
  [astro-ph/9512139].


  \bibitem{Calabrese:2013jyk} 
  E.~Calabrese {\it et al.},
   {\color{rossoCP3} Cosmological parameters from pre-planck cosmic microwave background measurements},
  Phys.\ Rev.\ D {\bf 87}, no. 10, 103012 (2013)
  doi:10.1103/PhysRevD.87.103012
  [arXiv:1302.1841 [astro-ph.CO]].

  \bibitem{Louis:2016ahn} 
  T.~Louis {\it et al.} [ACTPol Collaboration],
   {\color{rossoCP3} The Atacama Cosmology Telescope: Two-season ACTPol spectra and parameters},
  JCAP {\bf 1706}, 031 (2017)
  doi:10.1088/1475-7516/2017/06/031
  [arXiv:1610.02360 [astro-ph.CO]].

\bibitem{Baumann:2018muz} 
  D.~Baumann,
  {\color{rossoCP3} Primordial cosmology},
  PoS TASI {\bf 2017}, 009 (2018)
  doi:10.22323/1.305.0009
  [arXiv:1807.03098 [hep-th]].

  
  
  


\bibitem{Cuesta:2014asa} 
  A.~J.~Cuesta, L.~Verde, A.~Riess and R.~Jimenez,
  {\color{rossoCP3} Calibrating the cosmic distance scale ladder: the role of the sound horizon scale and the local expansion rate as distance anchors},
  Mon.\ Not.\ Roy.\ Astron.\ Soc.\  {\bf 448}, no. 4, 3463 (2015)
  doi:10.1093/mnras/stv261
  [arXiv:1411.1094 [astro-ph.CO]].

  
\bibitem{Wong:2019kwg} 
  K.~C.~Wong {\it et al.},
   {\color{rossoCP3} H0LiCOW XIII: A 2.4\% measurement of $H_{0}$ from lensed quasars: $5.3\sigma$ tension between early and late-Universe probes},
  arXiv:1907.04869 [astro-ph.CO].


\bibitem{Abbott:2019yzh} 
  B.~P.~Abbott {\it et al.} [LIGO Scientific and Virgo Collaborations],
   {\color{rossoCP3} A gravitational-wave measurement of the Hubble constant following the second observing run of Advanced LIGO and Virgo},
  arXiv:1908.06060 [astro-ph.CO].

\bibitem{Verde:2019ivm} 
  L.~Verde, T.~Treu and A.~G.~Riess,
 {\color{rossoCP3}   Tensions between the early and the late universe},
  Nature Astronomy {\bf 3}, 891  (2019)
  doi:10.1038/s41550-019-0902-0
  [arXiv:1907.10625 [astro-ph.CO]].


\bibitem{Tanabashi:2018oca} 
  M.~Tanabashi {\it et al.} [Particle Data Group],
{\color{rossoCP3}  Review of Particle Physics},
  Phys.\ Rev.\ D {\bf 98}, no. 3, 030001 (2018).
  doi:10.1103/PhysRevD.98.030001


 \bibitem{Steigman:1977kc} 
  G.~Steigman, D.N.~Schramm and J.E.~Gunn,
    {\color{rossoCP3} Cosmological limits to the number of massive leptons},
  Phys.\ Lett.\ B {\bf 66}, 202 (1977).
 
\bibitem{Kolb:1990vq} 
  E.~W.~Kolb and M.~S.~Turner,
   {\color{rossoCP3} The Early Universe},
  Front.\ Phys.\  {\bf 69}, 1 (1990).

  
\bibitem{Mangano:2005cc} 
  G.~Mangano, G.~Miele, S.~Pastor, T.~Pinto, O.~Pisanti and P.~D.~Serpico,
   {\color{rossoCP3} Relic neutrino decoupling including flavor oscillations},
  Nucl.\ Phys.\ B {\bf 729}, 221 (2005)
  doi:10.1016/j.nuclphysb.2005.09.041
  [hep-ph/0506164].


\bibitem{Aver:2015iza} 
  E.~Aver, K.~A.~Olive and E.~D.~Skillman,
   {\color{rossoCP3} The effects of He I $\lambda$10830 on helium abundance determinations},
  JCAP {\bf 1507}, 011 (2015)
  doi:10.1088/1475-7516/2015/07/011
  [arXiv:1503.08146 [astro-ph.CO]].


\bibitem{Peimbert:2016bdg} 
  A.~Peimbert, M.~Peimbert and V.~Luridiana,
   {\color{rossoCP3} The primordial helium abundance and the number of neutrino families},
  Rev.\ Mex.\ Astron.\ Astrofis.\  {\bf 52}, 419 (2016)
  [arXiv:1608.02062 [astro-ph.CO]].


\bibitem{Cooke:2017cwo} 
  R.~J.~Cooke, M.~Pettini and C.~C.~Steidel,
  {\color{rossoCP3} One percent determination of the primordial deuterium abundance},
  Astrophys.\ J.\  {\bf 855}, no. 2, 102 (2018)
  doi:10.3847/1538-4357/aaab53
  [arXiv:1710.11129 [astro-ph.CO]].

\bibitem{Pisanti:2007hk} 
  O.~Pisanti, A.~Cirillo, S.~Esposito, F.~Iocco, G.~Mangano, G.~Miele and P.~D.~Serpico,
    {\color{rossoCP3} PArthENoPE: Public algorithm evaluating the nucleosynthesis of primordial elements},
  Comput.\ Phys.\ Commun.\  {\bf 178}, 956 (2008)
  doi:10.1016/j.cpc.2008.02.015
  [arXiv:0705.0290 [astro-ph]].


\bibitem{Marcucci:2015yla} 
  L.~E.~Marcucci, G.~Mangano, A.~Kievsky and M.~Viviani,
    {\color{rossoCP3} Implication of the proton-deuteron radiative capture for Big Bang Nucleosynthesis},
  Phys.\ Rev.\ Lett.\  {\bf 116}, no. 10, 102501 (2016)
  Erratum: [Phys.\ Rev.\ Lett.\  {\bf 117}, no. 4, 049901 (2016)]
  doi:10.1103/PhysRevLett.116.102501, 10.1103/PhysRevLett.117.049901
  [arXiv:1510.07877 [nucl-th]].
  
\bibitem{Izotov:2014fga} 
  Y.~I.~Izotov, T.~X.~Thuan and N.~G.~Guseva,
  {\color{rossoCP3} A new determination of the primordial He abundance
    using the He I $\lambda 10830~{\buildrel _{\circ} \over {\mathrm{A}}}$ emission line: cosmological implications},
  Mon.\ Not.\ Roy.\ Astron.\ Soc.\  {\bf 445}, no. 1, 778 (2014)
  doi:10.1093/mnras/stu1771
  [arXiv:1408.6953 [astro-ph.CO]].

\bibitem{Anchordoqui:2019yzc} 
  L.~A.~Anchordoqui and S.~E.~Perez Bergliaffa,
  {\color{rossoCP3} Hot thermal universe endowed with massive dark vector fields and the Hubble tension},
  Phys.\ Rev.\ D {\bf 100}, no. 12, 123525 (2019)
  doi:10.1103/PhysRevD.100.123525
  [arXiv:1910.05860 [astro-ph.CO]].

  
\bibitem{Vagnozzi:2019ezj} 
  S.~Vagnozzi,
   {\color{rossoCP3} New physics in light of the $H_0$ tension: an alternative view},
  arXiv:1907.07569 [astro-ph.CO].

\bibitem{Obied:2018sgi} 
  G.~Obied, H.~Ooguri, L.~Spodyneiko and C.~Vafa,
    {\color{rossoCP3} de Sitter Space and the Swampland},
  arXiv:1806.08362 [hep-th].

  
\bibitem{Agrawal:2018own} 
  P.~Agrawal, G.~Obied, P.~J.~Steinhardt and C.~Vafa,
     {\color{rossoCP3} On the cosmological implications of the string swampland},
  Phys.\ Lett.\ B {\bf 784}, 271 (2018)
  doi:10.1016/j.physletb.2018.07.040
  [arXiv:1806.09718 [hep-th]].

\bibitem{Raveri:2018ddi} 
  M.~Raveri, W.~Hu and S.~Sethi,
   {\color{rossoCP3} Swampland conjectures and late-time cosmology},
  Phys.\ Rev.\ D {\bf 99}, no. 8, 083518 (2019)
  doi:10.1103/PhysRevD.99.083518
  [arXiv:1812.10448 [hep-th]].

\bibitem{Colgain:2019joh} 
  E.~\'O~Colg\'ain and H.~Yavartanoo,
   {\color{rossoCP3} Testing the Swampland: $H_0$ tension},
  arXiv:1905.02555 [astro-ph.CO].
  
\bibitem{Agrawal:2019dlm} 
  P.~Agrawal, G.~Obied and C.~Vafa,
    {\color{rossoCP3}$H_0$ tension, Swampland conjectures and the epoch of fading dark matter},
  arXiv:1906.08261 [astro-ph.CO].

\bibitem{Anchordoqui:2019amx} 
  L.~A.~Anchordoqui, I.~Antoniadis, D.~L\"ust, J.~F.~Soriano and T.~R.~Taylor,
   {\color{rossoCP3} $H_0$ tension and the string swampland},
  Phys.\ Rev.\ D {\bf 101}, 083532 (2020)
  doi:10.1103/PhysRevD.101.083532
  [arXiv:1912.00242 [hep-th]].

\bibitem{Salvatelli:2014zta} 
  V.~Salvatelli, N.~Said, M.~Bruni, A.~Melchiorri and D.~Wands,
  {\color{rossoCP3} Indications of a late-time interaction in the dark sector},
  Phys.\ Rev.\ Lett.\  {\bf 113}, no. 18, 181301 (2014)
  doi:10.1103/PhysRevLett.113.181301
  [arXiv:1406.7297 [astro-ph.CO]].


\bibitem{Kumar:2017dnp} 
  S.~Kumar and R.~C.~Nunes,
   {\color{rossoCP3} Echo of interactions in the dark sector},
  Phys.\ Rev.\ D {\bf 96}, no. 10, 103511 (2017)
  doi:10.1103/PhysRevD.96.103511
  [arXiv:1702.02143 [astro-ph.CO]].



  
  
 \bibitem{DiValentino:2017iww} 
  E.~Di Valentino, A.~Melchiorri and O.~Mena,
  {\color{rossoCP3} Can interacting dark energy solve the $H_0$ tension?},
  Phys.\ Rev.\ D {\bf 96}, no. 4, 043503 (2017)
  doi:10.1103/PhysRevD.96.043503
  [arXiv:1704.08342 [astro-ph.CO]].
  

\bibitem{Yang:2018euj} 
  W.~Yang, S.~Pan, E.~Di Valentino, R.~C.~Nunes, S.~Vagnozzi and D.~F.~Mota,
    {\color{rossoCP3} Tale of stable interacting dark energy, observational signatures, and the $H_0$ tension},
  JCAP {\bf 1809}, 019 (2018)
  doi:10.1088/1475-7516/2018/09/019
  [arXiv:1805.08252 [astro-ph.CO]].



\bibitem{Kumar:2019wfs} 
  S.~Kumar, R.~C.~Nunes and S.~K.~Yadav,
    {\color{rossoCP3} Dark sector interaction: a remedy of the tensions between CMB and LSS data},
  Eur.\ Phys.\ J.\ C {\bf 79}, no. 7, 576 (2019)
  doi:10.1140/epjc/s10052-019-7087-7
  [arXiv:1903.04865 [astro-ph.CO]].

  
\bibitem{DiValentino:2019ffd} 
  E.~Di Valentino, A.~Melchiorri, O.~Mena and S.~Vagnozzi,
    {\color{rossoCP3} Interacting dark energy after the latest Planck, DES, and $H_0$ measurements: an excellent solution to the $H_0$ and cosmic shear tensions},
  arXiv:1908.04281 [astro-ph.CO].

\bibitem{DiValentino:2019jae} 
  E.~Di Valentino, A.~Melchiorri, O.~Mena and S.~Vagnozzi,
    {\color{rossoCP3} Nonminimal dark sector physics and cosmological tensions},
  Phys.\ Rev.\ D {\bf 101}, no. 6, 063502 (2020)
  doi:10.1103/PhysRevD.101.063502
  [arXiv:1910.09853 [astro-ph.CO]].

\bibitem{Benevento:2020fev} 
  G.~Benevento, W.~Hu and M.~Raveri,
   {\color{rossoCP3} Can late dark energy transitions raise the Hubble constant?},
  arXiv:2002.11707 [astro-ph.CO].
  

\bibitem{Blumenhagen:2005mu} 
  R.~Blumenhagen, M.~Cveti\v c, P.~Langacker and G.~Shiu,
    {\color{rossoCP3} Toward realistic intersecting D-brane models},
  Ann.\ Rev.\ Nucl.\ Part.\ Sci.\  {\bf 55}, 71 (2005)
  doi:10.1146/annurev.nucl.55.090704.151541
  [hep-th/0502005].


\bibitem{Blumenhagen:2006ci} 
  R.~Blumenhagen, B.~Kors, D.~L\"ust and S.~Stieberger,
   {\color{rossoCP3}  Four-dimensional string compactifications with D-branes, orientifolds and fluxes},
  Phys.\ Rept.\  {\bf 445}, 1 (2007)
  doi:10.1016/j.physrep.2007.04.003
  [hep-th/0610327].

 \bibitem{Anchordoqui:2011nh}
  L.~A.~Anchordoqui and H.~Goldberg,
    {\color{rossoCP3} Neutrino cosmology after WMAP 7-year data and LHC first $Z'$ bounds},
  Phys.\ Rev.\ Lett.\  {\bf 108}, 081805 (2012)
  doi:10.1103/PhysRevLett.108.081805
  [arXiv:1111.7264 [hep-ph]].

\bibitem{Anchordoqui:2012wt}
  L.~A.~Anchordoqui, I.~Antoniadis, H.~Goldberg, X.~Huang, D.~L\"ust, T.~R.~Taylor and B.~Vlcek,
   {\color{rossoCP3} LHC phenomenology and cosmology of string-inspired intersecting D-brane models},
  Phys.\ Rev.\ D {\bf 86}, 066004 (2012)
  doi:10.1103/PhysRevD.86.066004
  [arXiv:1206.2537 [hep-ph]].

  

\bibitem{Anchordoqui:2012qu} 
  L.~A.~Anchordoqui, H.~Goldberg and G.~Steigman,
   {\color{rossoCP3} Right-handed neutrinos as the dark radiation: Status and forecasts for the LHC},
  Phys.\ Lett.\ B {\bf 718}, 1162 (2013)
  doi:10.1016/j.physletb.2012.12.019
  [arXiv:1211.0186 [hep-ph]].

\bibitem{Cremades:2003qj} 
  D.~Cremades, L.~E.~Iba\~nez and F.~Marchesano,
    {\color{rossoCP3} Yukawa couplings in intersecting D-brane models},
  JHEP {\bf 0307}, 038 (2003)
  doi:10.1088/1126-6708/2003/07/038
  [hep-th/0302105].


\bibitem{Anchordoqui:2011eg} 
  L.~A.~Anchordoqui, I.~Antoniadis, H.~Goldberg, X.~Huang, D.~L\"ust and T.~R.~Taylor,
    {\color{rossoCP3} $Z'$-gauge bosons as harbingers of low mass strings},
  Phys.\ Rev.\ D {\bf 85}, 086003 (2012)
  doi:10.1103/PhysRevD.85.086003
  [arXiv:1107.4309 [hep-ph]].
  

  
\bibitem{Brust:2013xpv} 
  C.~Brust, D.~E.~Kaplan and M.~T.~Walters,
   {\color{rossoCP3} New light species and the CMB},
  JHEP {\bf 1312}, 058 (2013)
  doi:10.1007/JHEP12(2013)058
  [arXiv:1303.5379 [hep-ph]].

  
\bibitem{Anchordoqui:2014dpa} 
  L.~A.~Anchordoqui, H.~Goldberg, X.~Huang and B.~J.~Vlcek,
   {\color{rossoCP3} Reconciling BICEP2 and Planck results with right-handed Dirac neutrinos in the fundamental representation of grand unified $E_6$},
  JCAP {\bf 1406}, 042 (2014)
  doi:10.1088/1475-7516/2014/06/042
  [arXiv:1404.1825 [hep-ph]].

\bibitem{Bazavov:2009zn} 
  A.~Bazavov {\it et al.},
   {\color{rossoCP3} Equation of state and QCD transition at finite temperature},
  Phys.\ Rev.\ D {\bf 80}, 014504 (2009)
  doi:10.1103/PhysRevD.80.014504
  [arXiv:0903.4379 [hep-lat]].


\bibitem{Laine:2006cp} 
  M.~Laine and Y.~Schroder,
   {\color{rossoCP3} Quark mass thresholds in QCD thermodynamics},
  Phys.\ Rev.\ D {\bf 73}, 085009 (2006)
  doi:10.1103/PhysRevD.73.085009
  [hep-ph/0603048].



  

\bibitem{Steigman:2012nb} 
  G.~Steigman, B.~Dasgupta and J.~F.~Beacom,
   {\color{rossoCP3} Precise relic WIMP abundance and its impact on searches for dark matter annihilation},
  Phys.\ Rev.\ D {\bf 86}, 023506 (2012)
  doi:10.1103/PhysRevD.86.023506
  [arXiv:1204.3622 [hep-ph]].


\bibitem{CMS:2019tbu} 
  CMS Collaboration [CMS Collaboration],
   {\color{rossoCP3} Search for a narrow resonance in high-mass dilepton final states in proton-proton collisions using 140$~\mathrm{fb}^{-1}$ of data at $\sqrt{s}=13~\mathrm{TeV}$},
  CMS-PAS-EXO-19-019.


\bibitem{Aad:2019hjw} 
  G.~Aad {\it et al.} [ATLAS Collaboration],
   {\color{rossoCP3} Search for new resonances in mass distributions of jet pairs using 139 fb$^{-1}$ of $pp$ collisions at $\sqrt{s}=13$ TeV with the ATLAS detector},
  JHEP {\bf 2003}, 145 (2020)
  doi:10.1007/JHEP03(2020)145
  [arXiv:1910.08447 [hep-ex]].

  

\bibitem{Aad:2020kep} 
  G.~Aad {\it et al.} [ATLAS Collaboration],
   {\color{rossoCP3} Search for dijet resonances in events with an isolated charged lepton using $\sqrt{s} = 13$ TeV proton-proton collision data collected by the ATLAS detector},
  arXiv:2002.11325 [hep-ex].


  
  
\bibitem{Galli:2010it} 
  S.~Galli, M.~Martinelli, A.~Melchiorri, L.~Pagano, B.~D.~Sherwin and D.~N.~Spergel,
  {\color{rossoCP3} Constraining fundamental physics with future CMB experiments},
  Phys.\ Rev.\ D {\bf 82}, 123504 (2010)
  doi:10.1103/PhysRevD.82.123504
  [arXiv:1005.3808 [astro-ph.CO]].




  
\bibitem{Abazajian:2019eic} 
  K.~Abazajian {\it et al.},
   {\color{rossoCP3} CMB-S4 science case, reference design, and project plan},
  arXiv:1907.04473 [astro-ph.IM].




  
  
\end{thebibliography}
\end{document}